# ACCRETION DISKS AROUND YOUNG STARS: THE CRADLES OF PLANET FORMATION

**Dmitry A. Semenov** – Max Planck Institute for Astronomy – Heidelberg – Department of Chemistry – Ludwig Maximilian University – Munich – Germany

**Richard D. Teague** – Center for Astrophysics – Harvard and Smithsonian – Boston – USA

**Protoplanetary disks around young stars are the birth sights of planetary systems like our own. Disks represent the gaseous dusty matter left after the formation of their central stars. The mass and luminosity of the star, initial disk mass and angular momentum, and gas viscosity govern disk evolution and accretion. Protoplanetary disks are the cosmic nurseries where microscopic dust grains grow into pebbles, planetesimals, and planets.**

## Protoplanetary Disks as Analogues of Young Solar Systems

Protoplanetary disks are believed to be the environments where planets, their satellites as well as comets and asteroids are formed over a timescale of several million years (Myr). Diversity of protoplanetary disks allows us to better comprehend properties of and conditions in the protosolar nebula, out of which our own solar system has emerged [1–3]. Protoplanetary disks are found around young stars with masses between ~0.1 and 2.5 solar masses ($1M_{Sun} \approx 2 \times 10^{30}$ kg). These rotating disks are made of ~99% gas and ~1% dust (by mass) and have typical masses of ~ 0.01–0.1$M_{Sun}$. Disks have sizes up to ~1000 au (1 au is the average distance between Earth and the Sun). For reference, Pluto orbits the Sun at ~ 40 au distance, while the Oort cloud of comets is located at distances beyond 2,000-200,000 au.

Protoplanetary disks are a natural outcome of the star formation process. The ~10 Myr evolution of a protoplanetary disk can be divided into three main stages. The newly born star is surrounded by a slowly rotating envelope of material which, due to the conservation of angular moment, flattens out into a circumstellar disk. During the first ~0.1 – 0.5 Myr, the central star continues to grow in mass by accreting matter from the inner edge of the disk, which, in turn, is fed by the remains of the extended envelope. Secondly, when the external supply of matter from the envelope onto the disk comes to an end, the star reaches its final mass and the disk mass begins to decrease over a timescale of ~3 – 10 Myr. Sometime during this entire process, the onset of planet formation commences, however it is currently unclear when exactly it begins. During the clearing phase that lasts several Myr, the remaining disk is dispersed by the planet-disk dynamical interactions and photoevaporation by stellar radiation.

## Protoplanetary Disk Structure

Protoplanetary disks are accretion disks, which evolution was mathematically described in the "classical" theoretical studies [4–6]. Their fundamental physical properties are described by the conservation laws for energy and angular momentum. An accretion disk is a flat, rotating gaseous structure in orbital motion around the gravitational center (a star). Usually the central star is much more massive than the disk, and the orbital motion of the gas at a radius $R$ nearly follows the Kepler law: $V_K = GM_*/R$ (where $G$ is the gravitational constant and $M_*$ is the mass of the star). The departure from the perfect Keplerian rotation is due to the thermal gradient, which makes gas orbital velocities to be slightly sub-Keplerian.

Accretion is the loss of potential energy via frictional dissipation. Accretion leads to a net transport of the disk matter toward the gravitational center, while a minority of the matter carries angular momentum outside of the disk. Measured accretion rates from disks onto their central stars are low, $\sim 10^{-12}$ to $10^{-8}$ $M_{Sun}$/yr [7]. The frictional dissipation is provided by the viscous stresses. The molecular viscosity is too low to drive the disk accretion. Thus, other mechanisms like turbulence or disk winds have been invoked. The origin of disk turbulence or winds are not yet fully understood and likely caused by various (magneto)-hydrodynamical instabilities [8,9]. In massive, cold disks a gravitational instability can occur, driving efficient angular momentum transport and accretion via spiral arms.

Turbulent viscosity in disks is often parameterized by the so-called $\alpha$ parameter: $\nu = \alpha c_S H$, where $c_S$ is the thermal gas velocity (sound speed) and $H$ is the scale height of the disk (local disk extend) [5]. Observations and advanced dynamical models of protoplanetary disks point to a weak subsonic turbulence with $\alpha \sim 10^{-5} - 10^{-3}$ [10]. By knowing the $\alpha$ parameter, the time evolution of the gas density distribution in a geometrically thin disk as a function of radius can be computed by solving for a 1D non-linear diffusion equation [6, 11].

Protoplanetary disks are characterized by the strong gradients of physical conditions and chemical composition, with cold and dense midplanes, warm upper layers, and hot, tenuous atmosphere (Fig. 1). The thermal disk structure is determined by the dissipation of accretion energy in the inner disk region, and by the reprocessing of the stellar and interstellar radiation in the rest of the disk. The dust and gas temperatures are coupled via collisions in dense layers, at particle densities above $10^5 - 10^8$ cm$^{-3}$. The disk gas becomes hotter than the dust in the atmosphere, reaching temperatures >10,000 K. At intermediate heights, where dust grains begin to shield the disk matter from external high-energy radiation, the dust and gas temperatures become close to each other, 20 – 500 K. In the inner disk, midplanes temperatures can reach up to 1,500 K due to accretion heating, while outer midplane regions are cold, 10 – 20 K.

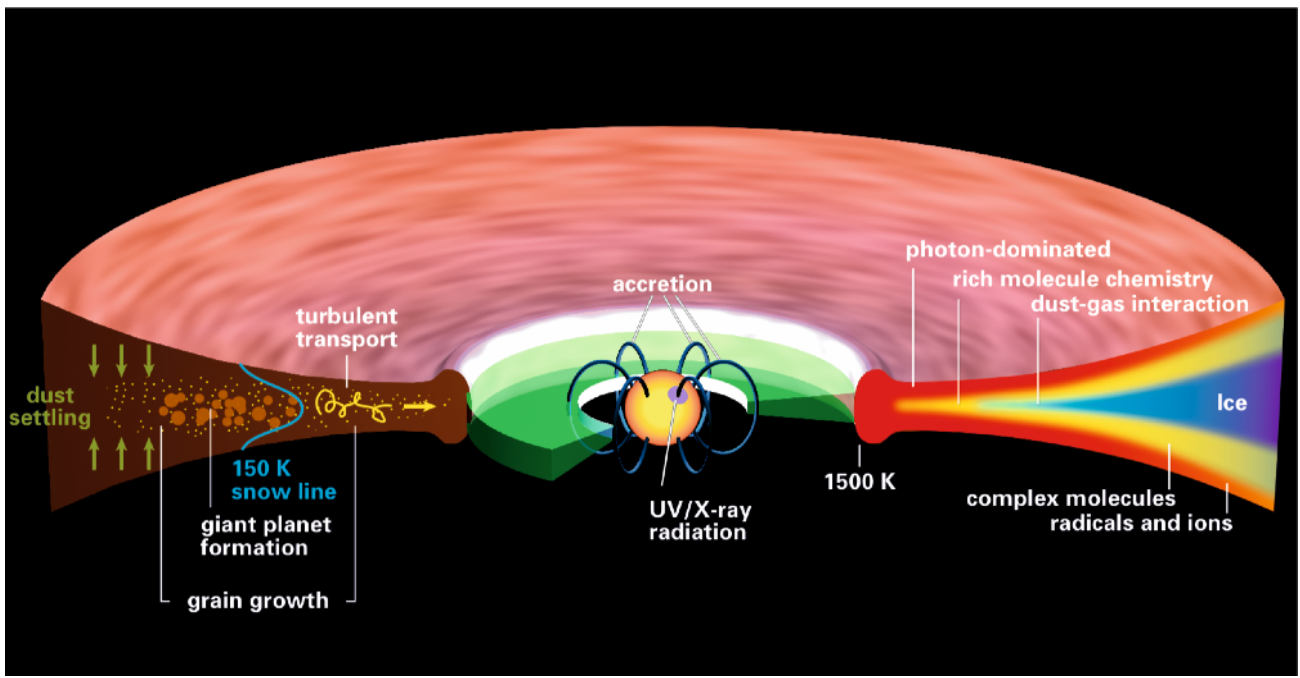

**Figure 1:** *A protoplanetary disk around a young Sun-like star. The young star illuminates the disk atmosphere, which leads to the flaring vertical structure and strong gradients of physical conditions. The µm-sized dust grains are coupled to the gas and provide opaqueness to the disk matter. The mm-sized dust grains gravitationally settle toward the disk midplane, where they can further grow and drift toward the inner regions. From chemical perspective, a disk can be divided into three layers: (1) a cold midplane where molecules are frozen onto dust grains as ices, (2) a warmer intermediate layer where molecules form and remain in the gas, and (3) a hot irradiated atmosphere with atomic species. Reprinted with permission from [2]. Copyright 2013 American Chemical Society.*

Disk gas density decreases very rapidly in the vertical direction. Assuming hydrostatic equilibrium between gravity and thermal pressure, the vertical gas density distribution can be approximated by a Gaussian function. The vertically integrated gas density also decreases radially outward. This often leads to a situation when more than a half of the disk mass is located inside the planet-forming zone at ≲ 20 – 50 au.

Dust evolution in disks leads to the dust density distribution that differs from the gas density distribution. Micron-sized dust grains remain collisionally coupled to the gas and follow the gas. They can grow through collisions into bigger millimeter-sized grains (pebbles), becoming partially decoupled from the gas and gravitationally settling towards the disk midplane. Pebbles experience a headwind from the gas due to the difference between their perfectly Keplerian velocities and the sub-Keplerian velocity of the gas. This leads to their rapid drift inward and loss, unless they are trapped between the disk gas gaps or grow far beyond 1 m in size. The most robust process for the dust growth beyond this 1 m barrier is the local accumulation of solids in turbulent eddies and other long-lived gas over-densities [12].

Later, these pebbles can grow into a swarm of km-sized planetesimals, first planetary embryos (~ 100 – 1000 km), and planetary cores (>1000 – 3000 km). According to the core-accretion scenario [13], giant planets like Jupiter can form when their solid cores grow above ~ 5 – 20 Earth masses and able to gravitationally attract nearby gas. Otherwise, the planetary cores end up as rocky, Earth-like planets. The planet formation timescale cannot be much longer than the disk lifetime of a ~10 Myr.

## Recent Developments

Recent progress in space-born and ground-based observatories like *ALMA*, *Herschel*, *NOEMA*, *Spitzer*, *VLT* provided images of hundreds of protoplanetary disks in thermal dust continuum and numerous molecular emission lines (see Figs. 2, 3). Micron-sized grains are best observed at visual and near-infrared wavelengths via scattered stellar light. Solid-state ice and silicate emission bands, and vibrational transitions of gaseous molecules are observed at near- and mid-infrared wavelengths (from ground and space). Bigger mm-sized grains and molecular rotational transitions are observed at (sub-)millimeter wavelengths with radio interferometers such as *ALMA*.

Surprisingly, a majority of disks observed at a high spatial resolution in dust continuum emission reveals a multitude of substructures like dark gaps, bright rings, spiral arms, etc., which are indicative of planet formation (Fig. 2). These circularly-symmetric gaps and rings are likely caused by embedded planets, while spirals could be induced by external perturbers. Ongoing large chemical surveys with *ALMA* and *NOEMA* will allow us to characterize the gas content in protoplanetary disks and to better understand what molecules are being delivered onto the atmospheres of young, forming planets. Recently, the first direct discovery of young, massive Jupiter-like planets in a few Myr old protoplanetary disk PDS 70 has been reported [16, 17].

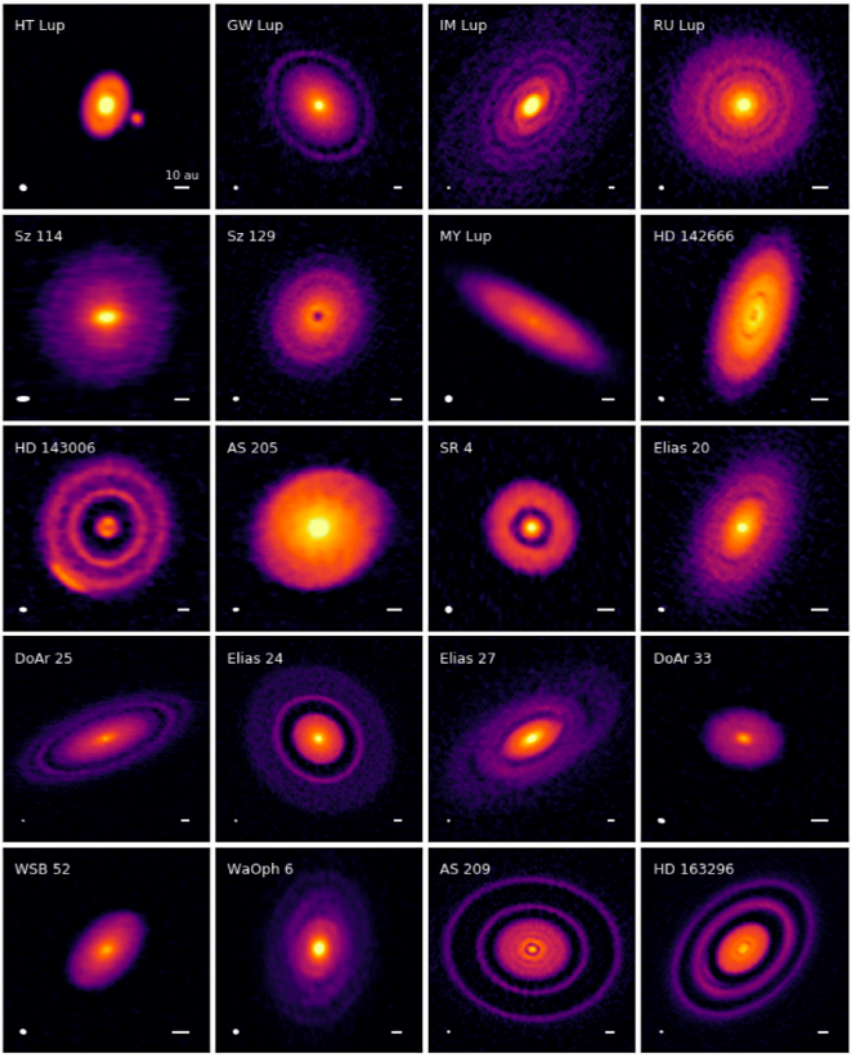

**Figure 2:** *Twenty protoplanetary disks observed at the wavelength of 1.25 mm with the Atacama Large Millimeter Array in Chile (DSHARP project; PI: Sean Andrews, Center for Astrophysics | Harvard and Smithsonian). What is visible in each panel is the continuum emission from ~1 mm-sized dust grains, which gravitationally settled toward the disk center. The detector pixel sizes (lower left corner) and the 10 au scale bars (lower right corners) are shown in each panel, respectively. The dark gaps are believed to be gravitationally cleared by giant planets. Reprinted with permission from [14]. Copyright 2018 Astrophysical Journal.*

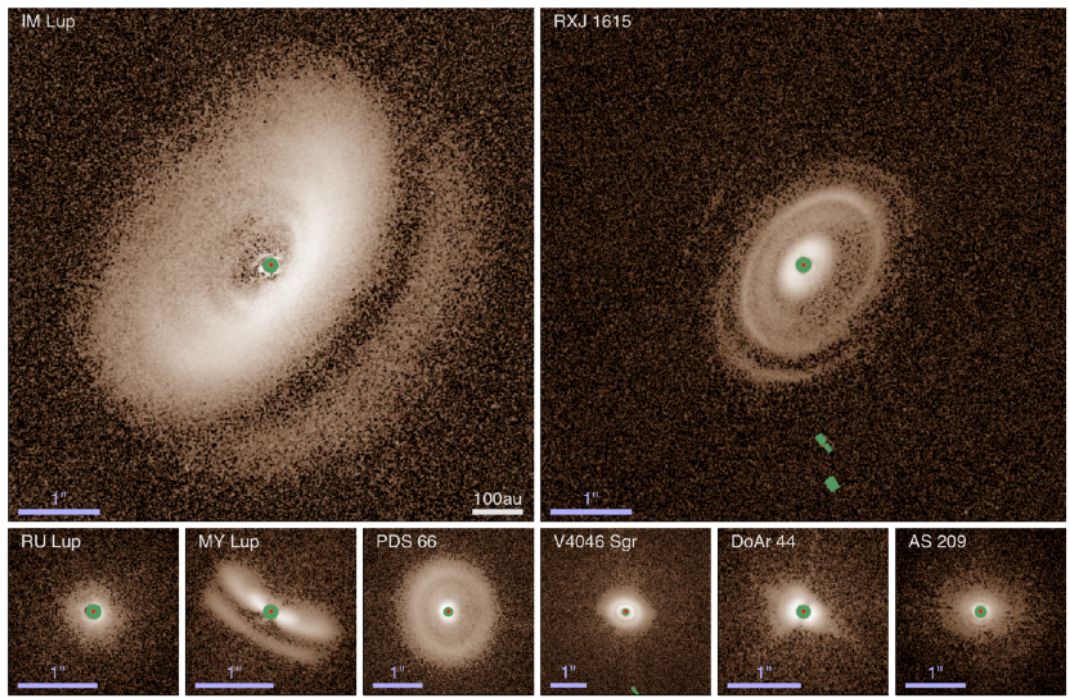

**Figure 3:** *Near-infrared images of eight protoplanetary disks (SPHERE instrument at the Very Large Telescope in Chile). What is visible in each panel is the stellar light scattered by sub-micron-sized dust grains. The strong emission of the central star is blocked by the coronographic mask. The dark lanes are the disk midplanes where dust absorbs the stellar light. The images are rescaled to represent the same physical size. Green areas mark places where no information is available. Reprinted with permission from [15]. Copyright 2018 Astrophysical Journal.*